\documentclass[showpacs,floatfix,preprint]{revtex4-1}
\usepackage{graphicx,psfrag,amsmath,amssymb,amsfonts,latexsym,color,epsf,graphpap,dcolumn}
\usepackage{physics,amsmath}

\date{today}
\usepackage[latin1]{inputenc}
\usepackage[T1]{fontenc} 
\usepackage[english]{babel}
\usepackage{bm}       
\usepackage{hyperref}
\pdfoutput=1

\begin{document}

\newcommand{\mn}{\mathbf n}
\newcommand{\mnk}{\mathbf {n k}}
\newcommand{\mm}{\mathbf m}
\newcommand{\mk}{\mathbf k}
\newcommand{\pp}{{\mathbf p}_\parallel}
\newcommand{\kp}{{\mathbf k}_\parallel}
\newcommand{\mv}[1]{\langle #1 \rangle}
\newcommand{\nk}[1]{#1^{(\mathbf n )}_{\mathbf k}}

\title{Stochastic particle creation: from the dynamical Casimir effect to cosmology}
\author{Mat\'ias Manti\~{n}an$^1$ \footnote{mantinanmatias@gmail.com}}
\author{Francisco D. Mazzitelli$^2$ \footnote{fdmazzi@cab.cnea.gov.ar}}

\author {Leonardo G. Trombetta$^3$
\footnote{trombetta@fzu.cz}}

\affiliation{$^1$ Department of Physics, University of Chicago, Chicago, IL 60637, USA }

\affiliation{$^2$ Centro At\'omico Bariloche and Instituto Balseiro,
Comisi\'on Nacional de Energ\'\i a At\'omica, 
R8402AGP Bariloche, Argentina}

\affiliation{$^3$ CEICO, Institute of Physics of the Czech Academy of Sciences, Na Slovance 1999/2, 182 21, Prague 8, Czechia}
\date{\today}
\begin{abstract}
We study a stochastic version of the dynamical Casimir effect, computing the particle creation inside a cavity  produced by a random motion of 
one of its walls. We first present a calculation  perturbative in the amplitude of the motion.  We compare the stochastic particle creation with the deterministic counterpart. Then we go beyond the perturbative evaluation using a stochastic version of the multiple scale analysis, that takes into account stochastic parametric resonance. We stress the relevance of the coupling between the different modes induced by the stochastic motion. In the single-mode approximation, the equations are formally analogous to those that describe the stochastic particle creation in a cosmological context, that we rederive using multiple scale analysis.
\end{abstract} 
\maketitle

\section{Introduction}
Quantum systems under the influence of time-dependent external conditions have attracted a lot of interest in the last decades. One of the interesting phenomena that arises in this context  is particle creation induced by time-dependent external backgrounds on quantum fields. The time dependence can be produced by time-dependent properties of the media in which the field propagates: by an external (e..g electromagnetic or gravitational) field, or by time-dependent boundary conditions.  Concrete examples are the so called dynamical Casimir effect (DCE) \cite{reviews}, in which photons are created by moving mirrors or by time-dependent electromagnetic properties, and gravitational particle creation for quantum field theories in curved spaces \cite{booksQFTCS}. The most studied situations in the last context are cosmological particle creation and Hawking radiation. 
Most previous works on the subject consider deterministic time-dependent external conditions. In this paper we are interested in the case of stochastic external conditions.

In the cosmological context, to our knowledge Ref. \cite{Hu:1997iu} addressed the problem for the first time,
considering a quantum field in the presence of a background metric with a random component. In 
more recent works 
\cite{Amin:2015ftc, Choudhury:2018rjl}, the 
stochasticity comes from the interaction of the quantum field with different environments. Technically, the dynamics of the system is analyzed using a correspondence to  one dimensional quantum mechanical problems with random impurities (more precisely, electrical resistance in wires with impurities). 
The role of noise in the early Universe has been discussed in \cite{Lozano:2020xga}.

Regarding previous works in the context of DCE, in Ref. \cite{Dodonov3} the authors considered a single harmonic oscillator and assumed that the time-dependent frequency consists of a  sequence  of barriers. The number of photons is evaluated assuming stochastic amplitudes  and stochastic separation between barriers. This separation turns to be more crucial than the amplitude fluctuations.  Contact is made with the problem of impurities in one dimensional chains, as in the cosmological approach described above. In Ref. \cite{Nori} the problem is treated at the quantum mechanical level for both the mirror and a single electromagnetic mode. On the other hand, in  Ref. \cite{Roman} the authors considered the DCE produced by a moving mirror that oscillates with a frequency given by $\Omega= 2 \omega_0 + K$, where 
$\omega_0$ is the lowest eigenfrequency of the modes in the cavity and $K$ is a stochastic detuning.

Different approaches have been used to tackle the problem of the deterministic DCE. 
In $1+1$ dimensions the conformal invariance allows for a simple solution, as described in the seminal paper by Moore \cite{Moore}. The problem can also be addressed using the so called instantaneous basis approach, in which the field is expanded in modes that satisfy the boundary conditions at each time \cite{Crocce2001}. This method can be extended to $3+1$ dimensions and also to the case of the full electromagnetic field. The equations for the modes become those of a set of coupled harmonic oscillators with time dependent frequencies and couplings.
To solve these dynamical equations one can work perturbatively in the amplitude of the motion of the walls. Alternatively, for oscillatory motion one can go beyond perturbation theory using the so called multiple scale analysis (MSA) \cite{Bender}. This method produces a resumation of the secular terms that appear in the perturbative calculations, and gives an exponential number of created particles due to parametric resonance. 

In this paper we discuss the phenomenon of stochastic particle creation (SPC). For the DCE, we go beyond the single-mode approximation discussed in previous works, computing the SPC first in a perturbative approach, and then using a stochastic version of the MSA. Due to parametric stochastic resonance, the number of created particles grows exponentially, although with a slower rate than in the deterministic case.  

For the case of a single mode, the dynamical equation that describes the DCE is equivalent to the equation of each mode of a scalar field in a cosmological context: a harmonic oscillator with stochastic frequency. We compare our results, based on the MSA, with those obtained in previous works.

The paper is organized as follows. In the next section we review the   standard approach to the DCE, based on the instantaneous basis approach for a quantum scalar field. 
In Section 3 we solve the equations for the modes assuming a stochastic motion of the wall. We compute the Bogoliubov coefficients perturbatively in the amplitude of the (random) motion, that depends on the correlation function of the noise.  We discuss whether it is possible or not to neglect the intermode couplings. In Section 4 we describe a nonperturbative calculation for a single mode, based on a stochastic generalization of the MSA method, as described originally in Ref.~\cite{Papanicolau}. 
We present a simpler version of the calculation, to highlight the relation between the stochastic and the deterministic MSA. 
In Section 5 we generalize the stochastic MSA for a set of coupled oscillators, and discuss its relevance in the DCE.
In Section 6 we use the results of Section 4 to analyze SPC in cosmology. Section 7  contains the conclusions of our work.


\section{Quantum scalar field in a cavity}

In this section we review  the quantization of a scalar field in a cavity, using the instantaneous basis formalism. We follow closely Ref. \cite{Crocce2001}, with slight changes of notation.

We consider a scalar field in a rectangular cavity formed by perfect mirrors, with dimensions $L_{x},L_{y},$ and $L_{z}$. The mirrors placed at $x=L_{x}$ and $y=L_y$ are at rest, while the other follows a prescribed trajectory
$z=L_{z}(t)$. We will assume that the motion starts at $t=0$ and finishes at $t=T$, when the mirror returns to its original position.

The scalar field $\phi(\mathbf{x},\mathit{t})$ satisfies the wave equation 
$\Box\phi=0$, and Dirichlet boundary conditions 
$\phi|_{\rm walls}=0$ for all times.
As usual, we expand the field
in terms of creation and annihilation operators
\begin{equation}
\phi(\mathbf{x},\mathit{t})=\sum_{\mathbf{n}}\left[
\hat{a}_{\mathbf{n}}^{\scriptscriptstyle{\rm in}}
u_{\mathbf{n}}(\mathbf{x},\mathit{t}) +\hat{a}_{\mathbf{n}}^{\dagger\, \scriptscriptstyle{\rm in}}
u_{\mathbf{n}}^*(\mathbf{x},\mathit{t}) \right] ,
\label{field}
\end{equation} 
where the mode functions $u_{\mathbf{n}}(\mathbf{x},\mathit{t})$ form
a complete orthonormal  set of solutions of the wave equation 
with vanishing boundary conditions. 

For $t\leq0$ the cavity is static and each field mode is 
given by
\begin{equation}
u_{\mathbf{n}}(\mathbf{x},\mathit{t})=\frac{1}{\sqrt{2\omega_
{\mathbf{n}}}}
\sqrt{\frac{2}{L_{x}}}\sin\left(\frac{n_{x}\pi}
{L_{x}} x\right)\sqrt{\frac{2}{L_{y}}}
\sin\left(\frac{n_{y}\pi}{L_{y}} y\right)
\sqrt{\frac{2}{L_{z}}}\sin\left(\frac{n_{z}\pi}{L_{z}} z\right)
e^{-i\omega_{\mathbf{n}}t} ,
\label{expest}
\end{equation}
with
\begin{equation}
\omega_{\mathbf{n}}=\pi\sqrt{\left(\frac{n_{x}}{L_{x}}\right)^{2} \!+ 
\left(\frac{n_{y}}{L_{y}}\right)^{2}\! + 
\left(\frac{n_{z}}{L_{z}}\right)^{2}} .
\label{omega}
\end{equation}
Here $n_{x},n_{y}$ and $n_{z}$ are positive integers and  $\mathbf{n}=(\mathit{n}_{x},\mathit{n}_{y},\mathit{n}_{z})$

In order to satisfy the boundary condition on the moving mirror, 
for $0<t<T$ we expand the mode 
functions in Eq.~\eqref{field} using an "instantaneous basis"
\begin{eqnarray}
u_{\mathbf{n}}(\mathbf{x},\mathit{t})&=&\sum_{\mathbf{k}}
Q_{\mathbf{k}}^{(\mathbf{n})}(t)\sqrt{\frac{2}{L_{z}(t)}}
\sin\left(\frac{k_{z}\pi}{L_{z}(t)} z\right)
\sqrt{\frac{2}{L_{y}}}\sin\left(\frac{k_{y}\pi}{L_{y}} y\right)
\sqrt{\frac{2}{L_{x}}}\sin\left(\frac{k_{x}\pi}{L_{x}} x\right)\\
&=&\sum_{\mathbf{k}}Q_{\mathbf{k}}^{(\mathbf{n})}(t)\,
\varphi_{\mathbf{k}}(\mathbf{x},\mathit{L}_{z}(t)) .
\label{exp}
\end{eqnarray}
The field degrees of freedom are now the functions $Q_{\mathbf{k}}^{(\mathbf{n})}(t)$, that satisfy the initial conditions
\begin{equation}
Q_{\mathbf{k}}^{(\mathbf{n})}(0)=\frac{1}{\sqrt{2\omega_{\mathbf{n}}}}\,
\delta_{\mathbf{k},\mathbf{n}}  \ ,\ \ 
\dot{Q}_{\mathbf{k}}^{(\mathbf{n})}(0)=-i
\sqrt{\frac{\omega_{\mathbf{n}}}{2}}\,\delta_{\mathbf{k},\mathbf{n}} .
\end{equation}

The field modes $u_{\mathbf{n}}(\mathbf{x},\mathit{t})$   must satisfy the 
wave equation. Inserting the expansion \eqref{exp} into the Klein Gordon equation, and taking into account that the $\varphi_{\mathbf{k}}$'s 
form a complete and orthonormal set, one can  obtain the following set of coupled 
equations for $Q_{\mathbf{k}}^{(\mathbf{n})}(t)$\cite{Crocce2001}:
\begin{equation}
\ddot{Q}_{\mathbf{k}}^{(\mathbf{n})}+\omega_{\mathbf{k}}^{2}(t)\,
Q_{\mathbf{k}}^{(\mathbf{n})}=2\lambda(t)\sum_{\mathbf{j}}
g_{\mathbf{kj}}\,\dot{Q}_{\mathbf{j}}^{(\mathbf{n})}+\dot{\lambda}(t)
\sum_{\mathbf{j}}g_{\mathbf{kj}}\,Q_{\mathbf{j}}^{(\mathbf{n})}+
\lambda^{2}(t)\sum_{\mathbf{j,l}}g_{\mathbf{lk}}\,g_{\mathbf{lj}}\,
Q_{\mathbf{j}}^{(\mathbf{n})} ,
\label{ecacop}
\end{equation}
where
\begin{equation}
\omega_{\mathbf{k}}(t)=\pi\sqrt{\left(\frac{k_{x}}{L_{x}}\right)^{2} \,+ 
\left(\frac{k_{y}}{L_{y}}\right)^{2}\, + \left(\frac{k_{z}}{L_{z}(t)}\right)^{2}}
\ \ ; \ \ \lambda(t)=\frac{\dot{L}_{z}(t)}{L_{z}(t)} .
\end{equation}
The coefficients $g_{\mathbf{kj}}$ are given by
\begin{equation}
g_{\mathbf{kj}}=L_{z}(t)\int_{0}^{L_{z}(t)}dz\ 
\frac{\partial\varphi_{\mathbf{k}}}{\partial L_{z}}\,\varphi_{\mathbf{j}},
\end{equation}
and can be explicitly computed:
\begin{equation}
g_{\mathbf{kj}}=-g_{\mathbf{jk}}= \left\{ 
\begin{array}{ll}
(-1)^{k_{z}+j_{z}}\frac{2k_{z}j_{z}}{j_{z}^{2}-k_{z}^{2}}\,
\delta_{k_{x}j_{x}}\,\delta_{k_{y}j_{y}} \,& \mbox{if $k_{z}\neq j_{z}$,} \\ 
0 & \mbox{if $k_{z}=j_{z}$}  .
\end{array}
\right.
\end{equation}

The annihilation and creation operators $\hat{a}_{\mathbf{k}}^
{\scriptscriptstyle{\rm in}}$ and $\hat{a}^{\dag\,\scriptscriptstyle{\rm in}}_
{\mathbf{k}}$ define the "in" vacuum state $\vert 0_{\rm{in}}\rangle$ and   the particle notion in the "in"
region ($t<0$). For $t>T$, when the wall stops,  we can introduce new 
operators, $\hat{a}_{\mathbf{k}}^{\scriptscriptstyle{\rm out}}$ and 
$\hat{a}^{\dag\,\scriptscriptstyle{\rm out}}_{\mathbf{k}}$, that define the "out" vacuum state $\vert 0_{\rm{out}}\rangle$ and   the particle notion in the "out" region. 

The "in" and "out" operators are connected by a Bogoliubov
transformation
\begin{equation}
\hat{a}_{\mathbf{k}}^{\scriptscriptstyle{\rm out}}=
\sum_{\mathbf{n}} ( \hat{a}_{\mathbf{n}}^{\scriptscriptstyle{\rm in}}\,
\alpha_{\mathbf{nk}}+\hat{a}^{\dag\,
\scriptscriptstyle{\rm in}}_{\mathbf{n}}\,\beta_{\mathbf{nk}}^{\star} ) .
\label{bog1}
\end{equation}
The coefficients $\alpha_{\mathbf{nk}}$ and $\beta_{\mathbf{nk}}$ 
can be obtained from the solutions of the coupled
equations \eqref{ecacop}. For $t>T$, the wall returns to its initial position and 
the solution reads
\begin{equation}
Q_{\mathbf{k}}^{(\mathbf{n})}(t)=
A_{\mathbf{k}}^{(\mathbf{n})}e^{-i\omega_{\mathbf{k}}t}+
B_{\mathbf{k}}^{(\mathbf{n})}e^{i\omega_{\mathbf{k}}t}\, .
\label{sol}
\end{equation}
The constant coefficients $A_{\mathbf{k}}^{(\mathbf{n})}$ and $B_{\mathbf{k}}^{(\mathbf{n})}$ 
are determined by the continuity of $Q_{\mathbf{k}}^{(\mathbf{n})}$ and $\dot Q_{\mathbf{k}}^{(\mathbf{n})}$ at $t=T$. 
Inserting Eq.~\eqref{sol} into Eqs.~\eqref{field} and \eqref{exp} we obtain 
an expansion of $\phi$ in terms of 
$\hat{a}_{\mathbf{k}}^{\scriptscriptstyle{\rm in}}$ and 
$\hat{a}_{\mathbf{k}}^{\dag\,\scriptscriptstyle{\rm in}}$, that is valid for
$t>T$. When this expression is compared with the equivalent expansion in terms of
$\hat{a}_{\mathbf{k}}^{\scriptscriptstyle{\rm out}}$ and 
$\hat{a}_{\mathbf{k}}
^{\dag\,\scriptscriptstyle{\rm out}}$,  one can see that
\begin{equation}
\alpha_{\mathbf{nk}}=\sqrt{2\omega_{\mathbf{k}}} 
A_{\mathbf{k}}^{(\mathbf{n})}\ \ , \ \ \beta_{\mathbf{nk}}=
\sqrt{2\omega_{\mathbf{k}}}\,B_{\mathbf{k}}^{(\mathbf{n})} .
\label{bog2}
\end{equation}

The number of particles created in the mode $\mathbf{k}$ is given by the mean  value 
of the number operator 
$\hat{a}_{\mathbf{k}}^{\dag\,\scriptscriptstyle{\rm out}}
\hat{a}_{\mathbf{k}}^{\scriptscriptstyle{\rm out}}$ in the
initial vacuum state 
 $|0_{\scriptscriptstyle{\rm in}}\rangle$. Using
Eqs.~\eqref{bog1} and \eqref{bog2} we obtain
\begin{equation}
\langle 0_{\rm in}|{\mathcal{N}}_{\mathbf{k}} |0_{\rm in}\rangle=\langle 
0_{\scriptscriptstyle{\rm in}}\mid 
\hat{a}_{\mathbf{k}}^{\dag\,\scriptscriptstyle{\rm out}} 
\hat{a}_{\mathbf{k}}^{\scriptscriptstyle{\rm out}}
\mid 0_{\scriptscriptstyle{\rm in}} \rangle = 
\sum_{\mathbf{n}}2\omega_{\mathbf{k}}|B_{\mathbf{k}}^{(\mathbf{n})}|^{2}=\sum_{\mathbf{n}}|\beta_{\mathbf{nk}}|^2
\label{numerodefotones} .
\end{equation}

All this formalism can be easily generalized for cylindrical  cavities of arbitrary section in the $\mathbf{x_\perp}=(x,y)$ plane.  In this case, the instantaneous basis reads
\begin{equation}
u_{\mathbf{n}}(\mathbf{x},\mathit{t})=\sum_{k_z, \mathbf{k_\perp}}
Q_{\mathbf{k}}^{(\mathbf{n})}(t)\sqrt{\frac{2}{L_{z}(t)}}
\sin\left(\frac{k_{z}\pi}{L_{z}(t)} z\right)
v_{\mathbf{k_\perp}}(\mathbf{x_\perp})\, ,
\end{equation}
with $\nabla^2 v_{\mathbf{k_\perp}}=-k_\perp^2 \, v_{\mathbf{k_\perp}}$. For the scalar field considered in this paper, the transversal functions $v_{\mathbf{k_\perp}}$ satisfy  Dirichlet boundary conditions on the lateral surface of the cylinder. When describing the electromagnetic field in terms of Hertz potentials, the transverse electric potential is a scalar field that satisfies Neumann boundary conditions on the lateral surface,  and Dirichlet boundary conditions on $z=0,\, z=L_z(t)$ \cite{Hertz}.   In any case, it is interesting to remark that the coupling coefficients
$g_{\mathbf{kj}}$ do not depend neither on the shape of the section of the cylinder nor on the boundary conditions on the lateral surface.


\section{Perturbative evaluation of the stochastic particle creation}

In this section we will solve Eqs.~\eqref{ecacop} perturbatively. We begin by writing these equations for a cavity of size $L_z(t) = L_0 (1 + \epsilon \xi(t))$,
\begin{equation}
\label{eq:modes_perturbative}
    \ddot{Q}_{\mathbf{k}}^{(\mathbf{n})} + \omega_{\mathbf{k}}^2 Q_{\mathbf{k}}^{(\mathbf{n})} = 2 \epsilon \xi(t) \omega_{k_z}^2 Q_{\mathbf{k}}^{(\mathbf{n})} + 2 \epsilon \dot{\xi}(t) \sum_{\mathbf m} g_{\mathbf{k m }}  \dot{Q}_{\mathbf{m}}^{(\mathbf{n})} +  \epsilon \ddot{\xi}(t) \sum_{\mathbf m} g_{\mathbf{k m }}  Q_{\mathbf{m}}^{(\mathbf{n})}\, ,
\end{equation}
where  $\omega_{k_z}^2=k_z^2\pi^2/L_0^2$.

In order to use perturbation theory, we propose a solution of the form
\begin{equation}
    Q_{\mathbf{k}}^{(\mathbf{n})}(t) = Q_{\mathbf{k}}^{(\mathbf{n})(0)}(t) + \epsilon Q_{\mathbf{k}}^{(\mathbf{n})(1)}(t) ,
\end{equation}
and plug it into Eq.~\eqref{eq:modes_perturbative}. At zeroth order in $\epsilon$ we have a simple harmonic oscillator, and the solution is
\begin{equation}
    Q_{\mathbf{k}}^{(\mathbf{n})(0)}(t) = \nk{A} e^{-i \omega_{\mathbf{k}} t} +\nk{B} e^{i \omega_{\mathbf{k}} t}, 
\end{equation}
where  $A_{\mathbf{k}}^{(\mathbf{n})}$ and $B_{\mathbf{k}}^{(\mathbf{n})}$ are  constants that will be fixed by the initial conditions. At first order we have
\begin{equation}
    \label{eq:modes_perturbative_order_1}
    \ddot{Q}_{\mathbf{k}}^{(\mathbf{n})(1)} + \omega_{\mathbf{k}}^2 Q_{\mathbf{k}}^{(\mathbf{n})(1)} = 2 \xi(t) \omega_{k_z}^2 Q_{\mathbf{k}}^{(\mathbf{n})(0)} + 2 \dot{\xi}(t) \sum_{\mathbf m} g_{\mathbf{k m }}  \dot{Q}_{\mathbf{m}}^{(\mathbf{n})(0)} +  \ddot{\xi}(t) \sum_{\mathbf m} g_{\mathbf{k m }}  Q_{\mathbf{m}}^{(\mathbf{n})(0)}.
\end{equation}
Eq.~\eqref{eq:modes_perturbative_order_1} corresponds to a driven harmonic oscillator, and can be solved using the corresponding Green's function
\begin{equation}
\begin{split}
    Q_{\mathbf{k}}^{(\mathbf{n})(1)}(t) &= \nk{C} e^{i \omega_{\mathbf{k}} t} + \nk{D} e^{-i \omega_{\mathbf{k}} t} + \int_0^\infty dt' \Theta(t-t') \frac{\sin[\omega_{\mathbf{k}} (t-t')]}{\omega_{\mathbf{k}}} \\ 
    &  \left[ 2 \xi(t') \omega_{k_z}^2 Q_{\mathbf{k}}^{(\mathbf{n})(0)}(t') + 2 \dot{\xi}(t') \sum_{\mathbf m} g_{\mathbf{k m }}  \dot{Q}_{\mathbf{m}}^{(\mathbf{n})(0)}(t') +  \ddot{\xi}(t) \sum_{\mathbf m} g_{\mathbf{k m }}  Q_{\mathbf{m}}^{(\mathbf{n})(0)}(t') \right]\, ,
\end{split}
\end{equation}
where we added an homogeneous solution with arbitrary constants $\nk{C}$  and $\nk{D} $.

Replacing the result for $Q_{\mathbf{k}}^{(\mathbf{n})(0)}$ and performing some integrations by parts, we find, for $t\geq T$, 
\begin{equation}
\begin{split}
    Q_{\mathbf{k}}^{(\mathbf{n})(1)}(t) &= \nk{C} e^{-i \omega_{\mathbf{k}} t} + \nk{D} e^{i \omega_{\mathbf{k}} t} + 2 \omega_{\mathbf{k}} \int_0^T dt' \xi(t') \sin[\omega_{\mathbf{k}} (t-t')] \\ 
    &  \sum_{\mathbf m} \left(  \nk{A} e^{-i \omega_{\mathbf{k}} t'} + \nk{B} e^{i \omega_{\mathbf{k}} t'} \right) \left( \delta_{\mathbf {m k }}\frac{\omega_{k_z}^2}{\omega_{\mathbf k}^2} + g_{\mathbf{k m }} \frac{\omega_{\mathbf{m}}^2-\omega_{\mathbf{k}}^2}{2 \omega_{\mathbf{k}}^2} \right).
\end{split}
\end{equation}
After imposing the initial conditions  $\nk{B}= \nk{C}=\nk{D}=0$ and $\nk{A} = \frac{\delta_{\mathbf {n,k }}}{\sqrt{2 \omega_{\mathbf{k}}}}$, we get
\begin{equation}
\begin{split}
    Q_{\mathbf{k}}^{(\mathbf{n})}(t) &= \frac{\delta_{\mathbf{k n}}}{\sqrt{2 \omega_{\mathbf{k}}}} e^{-i \omega_{\mathbf{k}} t} + \epsilon \frac{\sqrt{2 \omega_{\mathbf{k}}}}{\omega_{\mathbf{k}}}v_{\mathbf{n k}} \int_0^T dt' \xi(t') \sin[\omega_{\mathbf{k}} (t-t')] e^{-i \omega_{\mathbf{k}} t'},
\end{split}
\end{equation}
where
\begin{equation}
    v_{\mathbf{n k}} =  \delta_{\mathbf {n k }} \frac{\omega_{k_z}^2}{\omega_{\mathbf{k}}}+ g_{\mathbf{k n }} \frac{\omega_{\mathbf{n}}^2-\omega_{\mathbf{k}}^2}{2 \sqrt{\omega_{\mathbf{k}} \omega_{\mathbf{n }}}} .
\end{equation}
The Bogoliubov coefficient can be obtained as follows
\begin{equation}
    \beta_{\mathbf{n k }} = \left( i \omega_{\mathbf{k}} Q_{\mathbf{k}}^{(\mathbf{n})}(t) + \dot{Q}_{\mathbf{k}}^{(\mathbf{n})}(t)  \right) \frac{e^{- i \omega_{\mathbf{k}} T}}{\sqrt{2 \omega_{\mathbf{k}}}}.
\end{equation}
Therefore, replacing our perturbative solution for $\dot{Q}_{\mathbf k}^{\mathbf (n)}$, we find
\begin{equation}
    \beta_{\mathbf{n k }} = - i \epsilon v_{\mathbf {n k}}\int_0^T dt'\, \xi(t') e^{-i (\omega_{\mathbf{n }} + \omega_{\mathbf{k}})t'} ,
\end{equation}
and 
\begin{equation}\label{Bog-gral}
    |\beta_{\mathbf{n k }}|^2 = \epsilon^2 v_{\mathbf{n k}}^2 \int_0^T \int_0^T dt' dt'' \xi(t')\xi(t'') e^{-i (\omega_{\mathbf{n }} + \omega_{\mathbf{k}})(t'-t'')}.
\end{equation}

So far, we have not imposed any condition on $\xi(t)$. Now, we study the case where we have an stochastic noise described by
\begin{equation}\label{eq:noisecf}
    \mv{\xi (t)} =0 \, ,\quad\quad
    \mv{\xi (t)\xi (t')} = R(t-t')\, ,
\end{equation}
where $R(t)$ is the noise correlation, and its Fourier transform is
\begin{equation}
\label{eq:fourier_correlation_definition}
    S(\nu) = \int_0^\infty R(u) e^{i \nu u}.
\end{equation}
For this particular choice of $\xi(t)$ we get
\begin{equation}\label{eq:betank_fin}
    \mv{|\beta_{\mathbf{n k }}|^2} = 2 \epsilon^2 T v_{\mathbf{n k}}^2 \Re[S(\omega_{\mathbf{n }} + \omega_{\mathbf{k}})],
\end{equation}
and therefore
\begin{equation}
\langle{\mathcal{N}}_{\mathbf{k}} \rangle=2 \epsilon^2 T \sum_{\mathbf n}
  v_{\mathbf{n k}}^2 \Re[S(\omega_{\mathbf{n }} + \omega_{\mathbf{k}})]\, .
\end{equation}
This is the main result of this section. Note that the intermode coupling can be neglected only when the spectrum of the noise is sufficiently peaked around  a single mode. 
To arrive to Eq.~\eqref{eq:betank_fin}, we assumed that $\omega T\gg 1$, where $\omega$ represents the frequencies of the relevant modes in the cavity.  However, the perturbative result will be valid as long as the higher order corrections are smaller than the leading order. and this will be the case when $\epsilon^2\omega T\lesssim 1$.

It is interesting to compare the SPC with the deterministic counterpart. To do this, we come back to Eq.~\eqref{Bog-gral} and assume that $\xi(t)$ is a deterministic function. We see that  $|\beta_{\mathbf{n k }}|^2$ is proportional to $\vert\widetilde\xi(\omega_\mk+\omega_\mn)\vert^2$, where $\widetilde\xi(\nu)$ is the Fourier transform of $\xi(t)$. In particular, for an oscillatory motion
$\xi(t)= \sin(\Omega t)$ we obtain
\begin{equation}
    |\beta_{\mathbf{n k }}|^2 = \frac{1}{4}\epsilon^2
v_{\mathbf{n k}}^2 T^2\delta_{\Omega,\omega_\mk+\omega_\mm}\, . 
\end{equation}
We see that, while for the stochastic motion the total number of created particles grows as $\epsilon^2 T$, for a deterministic, oscillatory motion,  it grows as $\epsilon^2 T^2$\cite{Ji}. 
These two time scales, $\epsilon T$ and $\epsilon^2 T$, will reappear in the nonperturbative calculations.

\section{Non perturbative calculation: single mode}
So far we have analysed the system using a perturbative approach. However, this approach fails to describe the system at large times. One solution is to use the MSA, also called two times method, which allows us to resum the essential contributions of a stochastic perturbation in order to get an uniformly valid solution \cite{Bender,Papanicolau}.

In order to illustrate the method, we will first obtain the solution to the equation of a harmonic oscillator $Q(t)$ with multiplicative noise (stochastic time-dependent frequency) using MSA. Then we will apply a similar approach to compute $|Q|^2$, that will allow us to obtain the squared modulus of the Bogoliubov coefficients.

Before doing this, we remind the reader that, for a harmonic oscillator with a deterministic time-dependent frequency,
\begin{equation}
    \ddot{Q} + \omega^2 [1 + \epsilon \sin(2 \omega t)]Q = 0,
\end{equation}
the naive perturbative expansion in powers of $\epsilon$ contains secular terms that are $O((\epsilon\omega t)^n)$, where $n$ is the order of the approximation. The perturbative solution is therefore valid at short times $\epsilon \omega t\ll 1$. To remediate this, one introduces a second time scale $\bar\tau=\epsilon t$
and expands the solution as
\begin{equation}
    Q(t) = Y^{(0)}(t,\bar\tau) + \epsilon Y^{(1)}(t,\bar\tau) + O(\epsilon^2).
\end{equation}
The functions $Y^{(0)}$ and $Y^{(1)}$ satisfy
\begin{eqnarray}\label{eq:basicMSA}
    &&\partial_t^2 Y^{(0)} + \omega^2 Y^{(0)} = 0, \nonumber\\
    && \partial_t^2 Y^{(1)} + \omega^2 Y^{(1)} = - \omega^2 \sin(2\omega t) Y^{(0)} - 2\partial^2_{t\bar\tau} Y^{(0)} \, .
    \end{eqnarray}
The leading order solution reads
\begin{equation}
    Y^{(0)}(t,\bar\tau) = A(\bar\tau) e^{-i\omega t} + B(\bar\tau) e^{i\omega t},
\end{equation}
and the functions $A(\bar\tau)$ and $B(\bar\tau)$ are chosen such that there are no secular terms in $Y^{(1)}(t,\bar\tau)$, that is, no terms proportional to $e^{\pm i \omega t}$ in the r.h.s. of Eq.~\eqref{eq:basicMSA}. A simple calculation gives
\begin{equation}
\frac{dA}{d\bar\tau}=\frac{\omega}{4} B\, ,\quad\quad
\frac{dB}{d\bar\tau}=\frac{\omega}{4} A\, .
\end{equation}
Therefore, 
with the initial conditions $A(0)=1/\sqrt{2\omega}\, , B(0)=0$ we obtain
\begin{equation}
A(\bar\tau)=\frac{1}{\sqrt{2\omega}}\cosh\left(\frac{\omega}{4}\bar\tau \right)\, ,\quad\quad
B(\bar\tau)=\frac{1}{\sqrt{2\omega}}\sinh\left(\frac{\omega}{4}\bar\tau \right)\, ,
\end{equation}
and
\begin{equation}\label{eq:Bog_single_det}
|\beta|^2 = \sinh^2\left(\frac{\omega}{4}\bar\tau\right)= |\alpha|^2-1\, .
\end{equation}
This shows the amplification of the oscillations due to parametric resonance. We turn now to the stochastic case.

\subsection{MSA for a stochastic  harmonic oscillator}
We will start our discussion of the stochastic MSA by studying a simple example,   a random harmonic oscillator, given by
\begin{equation}
\label{eq:random_harmonic_oscillator}
    \ddot{Q} + \omega^2 [1 + \epsilon \xi(t)]Q = 0,
\end{equation}
with initial conditions
\begin{equation}\label{icQ}
    Q(0) = \frac{1}{\sqrt{2\omega}}, \,\,\,\,
    \dot{Q}(0) = -i\sqrt{\frac{\omega}{2}},
\end{equation}
where the frequency has a stochastic component given by $\xi(t)$. A similar problem can be found in \cite{Papanicolau}. 
As before, the noise is characterized by its mean value and its correlation function $R(t-t')$, see Eq.~\eqref{eq:noisecf}.

Following the MSA method \cite{Crocce2001,Bender,Papanicolau}, we introduce the new time scale $\tau = \epsilon^2 t$,  
and expand our solution in powers of $\epsilon$ as follows
\begin{equation}\label{eq:expQ}
    Q(t) = Y^{(0)}(t,\tau) + \epsilon Y^{(1)}(t,\tau) + \epsilon^2 Y^{(2)}(t,\tau) + O(\epsilon^3).
\end{equation}
Introducing this expansion in Eq.~\eqref{eq:random_harmonic_oscillator} we get the set of equations
\begin{eqnarray}
\label{eq:rho_perturbation_equations}
    &&\partial_t^2 Y^{(0)} + \omega^2 Y^{(0)} = 0, \nonumber\\
    && \partial_t^2 Y^{(1)} + \omega^2 Y^{(1)} + \omega^2 \xi(t) Y^{(0)} = 0,\nonumber \\
   && \partial_t^2 Y^{(2)} + \omega^2 Y^{(2)} + \omega^2 \xi(t) Y^{(1)} + 2 \partial_{\tau t}^2 Y^{(0)} = 0.
\end{eqnarray}
First, we notice that $Y^{(0)}$ satisfies the equation of a simple harmonic oscillator of frequency $\omega$, so the solution to the first equation is 
\begin{equation}
\label{eq:rho_sol_order0}
    Y^{(0)}(t,\tau) = A(\tau) e^{-i\omega t} + B(\tau) e^{i\omega t},
\end{equation}
where $A(\tau)$ and $B(\tau)$ are the slowly varying functions we want to determine. These functions will be fixed by imposing that the stochastic mean values $\mv{Y^{(1)}}$ and $\mv{Y^{(2)}}$ do not have secular terms.

The second equation in \eqref{eq:rho_perturbation_equations} is that of a harmonic oscillator with a source $-\omega^2 \xi(t) Y^{(0)} $. To solve this equation we use the harmonic oscillator propagator to compute the contribution of the source, and we use the solution \eqref{eq:rho_sol_order0}, obtaining
\begin{equation}\label{eq:solY1}
    Y^{(1)}(t,\tau) = C(\tau) e^{-i\omega t} + D(\tau) e^{i\omega t} - \omega \int_0^t dt' \sin[\omega(t-t')] \xi(t') \left[ A(\tau) e^{-i\omega t'} + B(\tau) e^{i\omega t'} \right],
\end{equation}
 where $C(\tau)$ and $D(\tau)$ are arbitary functions. Note that the $\tau$ under the integral sign is not being integrated since we are solving partial differential equations where $\tau$ and $t$ are independent variables. Furthermore, averaging over the noise we find that
\begin{equation}
    \mv{Y^{(1)}(t,\tau)} = C(\tau) e^{-i\omega t} + D(\tau) e^{i\omega t}.
\end{equation}
Being linear in the noise, the source has vanishing mean value
and $\mv {Y^{(1)}}$ has no secular terms. Therefore, there are no conditions imposed on $A(\tau)$ or $B(\tau)$ yet. As a consequence, it is necessary to consider the terms of order $O(\epsilon^2)$. This is also the reason why the new timescale for a  stochastic  frequency is $\tau=\epsilon^2 t$ and not $\bar\tau=\epsilon t$,  as in the deterministic case. 

Solving the last equation in \eqref{eq:rho_perturbation_equations} by the same method we get
\begin{equation}
\begin{split}
    Y^{(2)}(t,\tau) = &E(\tau) e^{-i\omega t} + F(\tau) e^{i\omega t} + 2 i \int_0^t dt' \sin[\omega(t-t')] \left[ A'(\tau) e^{-i\omega t'} - B'(\tau) e^{i\omega t'} \right] \\
    & -\omega \int_0^t dt' \sin[\omega(t-t')] \xi(t') \bigg\{ C(\tau) e^{-i\omega t'} + D(\tau) e^{i\omega t'}  \\ 
    & -\omega \int_0^{t'} dt'' \sin[\omega(t'-t'')] \xi(t'') \left[ A(\tau) e^{-i\omega t''} + B(\tau) e^{i\omega t''} \right] \bigg\} ,
\end{split}
\end{equation}
where a prime denotes derivative with respect to $\tau$,  and 
$E(\tau)$ and $F(\tau)$ are arbitrary functions.
Now, if we take the mean value of $Y^{(2)}$ we obtain
\begin{equation}\label{eq:mvY2}
\begin{split}
    \mv{Y^{(2)}(t,\tau)} = &E(\tau) e^{-i\omega t} + F(\tau) e^{i\omega t} + \int_0^t dt' \sin[\omega (t-t')] \bigg\{ 2 i \left[ A'(\tau) e^{-i\omega t'} - B'(\tau) e^{i\omega t'} \right] \\
    & +\omega^2  \int_0^{t'} dt'' \sin[\omega(t'-t'')] R(t'-t'') \left[ A(\tau) e^{-i\omega t''} + B(\tau) e^{i\omega t''} \right] \bigg\},
\end{split}
\end{equation}
that can be rewritten as
\begin{equation}\label{eq:mvY2bis}
\begin{split}
    \mv{Y^{(2)}(t,\tau)} = &E(\tau) e^{-i\omega t} + F(\tau) e^{i\omega t} + \int_0^t dt' \sin[\omega (t-t')] \bigg\{ 2 i \left[ A'(\tau) e^{-i\omega t'} - B'(\tau) e^{i\omega t'} \right] \\
    & +\omega^2  \int_0^{t'} du \sin[\omega u] R(u) \left[ A(\tau) e^{-i\omega (t'-u)} + B(\tau) e^{i\omega (t'-u)} \right] \bigg\}\, .
\end{split}
\end{equation}
We notice that there are secular terms, i.e. terms proportional to $e^{\pm i \omega t'}$ inside the $t'$-integral. So we choose $A(\tau)$ and $B(\tau)$ such that these secular terms do vanish. Therefore, we set
\begin{equation}
\begin{split}
    0 = &e^{-i\omega t'} \left[ A'(\tau) - A(\tau) \frac{\omega ^2}{4} \int_0^{t'} du R(u) - B(\tau) e^{2i\omega t'} \int_0^{t'} du R(u) e^{-2 i \omega u} \right] \\
    &+ e^{i\omega t'} \left[ B'(\tau) + B(\tau) \frac{\omega^2}{4} \int_0^{t'} du R(u) + A(\tau) e^{-2i\omega t'} \int_0^{t'} du R(u) e^{2 i \omega u} \right].
\end{split}
\end{equation}
In terms of the Fourier transform of the correlation Eq.~\eqref{eq:fourier_correlation_definition} 
we get, in the large $t'$ limit
\begin{eqnarray}
    A'(\tau) &=& \frac{\omega^2}{4} \left[S(2\omega) - S(0) \right] A(\tau)\nonumber \\
    B'(\tau) &=& \frac{\omega^2}{4} \left[S(2\omega)^* - S(0)  \right] B(\tau) .
\end{eqnarray}
Solving these equations with the initial conditions Eq.~\eqref{icQ}
we obtain
\begin{equation}
Q(t)=\frac{1}{\sqrt{2\omega}}e^{-i\omega t} e^{\frac{\omega^2}{4} \left[S(2\omega) - S(0) \right]\epsilon^2 t}+ O(\epsilon)\, .
\end{equation}
We see that the solution has an exponential growth   as long as 
$\Re[S(2\omega) - S(0)]>0$. This is a stochastic counterpart of the deterministic parametric resonance.
The noise also produces a shift in the frequency of oscillation:
$\omega\to \omega - \epsilon^2\omega^2\Im[S(2\omega)]/4$.

Notice that we only used the expansion in $\epsilon$ to find the conditions on the slowly varying functions $A(\tau)$ and $B(\tau)$ that ensure vanishing secular terms in the lowest order of the expansion. This is the main idea of the MSA. Moreover, the solution we have found,  $Q(t)\simeq Y^{(0)}$,  is  valid  to lowest order in $\epsilon$. However, as we know that 
$\mv{Y^{(1)}}=0$, 
the mean value of $Q(t)$ satisfies
\begin{equation}\label{mvQ}
\mv{Q(t)}=\frac{1}{\sqrt{2\omega}}e^{-i\omega t} e^{\frac{\omega^2}{4} \left[S(2\omega) - S(0) \right]\epsilon^2 t} + O(\epsilon^2)\, .
\end{equation}
Had we solved the stochastic differential equation for $Q(t)$ with the  initial conditions
\begin{equation}\label{icQ2}
    Q(0) = 1, \,\,\,\,
    \dot{Q}(0) = 0 \, ,
\end{equation}
we would have obtained
\begin{equation}
    \mv{Q(t)} = e^{\frac{\omega^2}{4} \left[ \Re[S(2\omega)] - S(0) \right] \epsilon^2 t} \cos\left[\left(\omega- \frac{\omega^2 \epsilon^2}{4} \Im[S(2\omega)]\right)t \right] + O(\epsilon^2) \, ,
\end{equation}
which coincides with the result obtained in Ref.~\cite{Papanicolau}
using a different approach.

\subsection{Bogoliubov coefficients using MSA}

If the random motion stops at $t=T$, for $t>T$ we will have
\begin{equation}
    Q(t)= \frac{\alpha}{\sqrt{2\omega}}e^{-i\omega t}+ \frac{\beta}{\sqrt{2\omega}}e^{i\omega t}\, ,
\end{equation}
where $\alpha$ and $\beta$ are the Bogoliubov coefficients. One could read the stochastic mean values of 
$\alpha$ and $\beta$ from Eq.~\eqref{mvQ}. However, the number of created particles is given by
$\mv{|\beta|^2}$ and, as $Q(t)$ is a stochastic variable, one cannot obtain  $\mv{|\beta|^2}$ from $\mv {Q(t)}$. 
Our strategy to compute $\mv{|\beta|^2}$ will be to implement the MSA imposing that  $\mv{ |Q(t)|^2}$ does  not include secular terms. Taking into account that, for $t>T$, 
\begin{equation}\label{eq:modQ2Bog}
 |Q(t)|^2 = \frac{1}{2 \omega}
 \left(|\alpha|^2 +|\beta|^2 + 
 2 \Re[\alpha\beta^* e^{-2 i \omega t}]\right)\, ,
\end{equation}
and that $|\alpha|^2 -|\beta|^2=1$, we will be able to compute $\mv{|\beta|^2}$ from the non-oscillating part of $\mv{ |Q(t)|^2}$. 

From Eq.~\eqref{eq:expQ} we have
\begin{equation}
    \mv{|Q(t)|^2}=|Y^{(0)}|^2+\epsilon^2\mv{2\Re[Y^{(0) *}Y^{(2)}]+|Y^{(1)}|^2}\,.
\end{equation}
In the previous section, we determined $A(\tau)$ and $B(\tau)$ from the absence of secular terms in $\mv{Y^{(2)}}$. Here, these functions will be such that no secular terms appear in 
\begin{equation}\label{eq:condmodY2}
2\Re[Y^{(0) *}\mv{Y^{(2)}}]+\mv {|Y^{(1)}|^2}.
\end{equation}
We have already computed $\mv{Y^{(2)}}$ in Eq.~\eqref{eq:mvY2bis}. The mean value  $\mv {|Y^{(1)}|^2}$ can be obtained from Eq.~\eqref{eq:solY1} and reads
\begin{eqnarray}
\label{eq:mvY12}
\mv{|Y^{(1)}|^2}&=&|C|^2+|D|^2+2\Re[C^*D e^{2 i \omega t}]+\omega^2\int_0^t dt' 
  \int_0^t \sin[\omega(t-t')]dt''\sin[\omega(t-t'')] \nonumber\\
&&\times R(t'-t'') (A e^{-i\omega t'} +B e^{i\omega t'} )(A^* e^{i\omega t''} +B^* e^{-i\omega t''} )\, .
\end{eqnarray}
In order to isolate the secular terms, it is useful to note that, for a function $f(t,t')$ such that $f(t',t'')=f(t'',t')^*$ we have
\begin{eqnarray}
&& \int_0^t dt' \int_0^t d t''\, f(t',t'')= 2 \Re\left[ \int_0^t  dt'\int_0^{t'} dt'' \, f(t',t'')\right]\nonumber\\
&&  2 \Re\left[ \int_0^t dt'\int_0^{t'} du\, f(t',t'-u)\right]\, .
\end{eqnarray}
Using this property in Eq.~\eqref{eq:mvY12}, inserting the result into
 Eq.~\eqref{eq:condmodY2} and using Eq.~\eqref{eq:mvY2bis} we find, after a straightforward calculation, the differential equations that must satisfy $A(\tau)$ and $B(\tau)$ to cancel the secular terms in $\mv{|Q(t)|^2}$. They are given by
\begin{eqnarray}\label{eq:diffAyB2}
&& (|A(\tau)|^2+|B(\tau)|^2)'=\omega^2\Re[S(2\omega)](|A(\tau)|^2+|B(\tau)|^2),\nonumber\\
&& (A^*(\tau)B(\tau))'=\frac{\omega^2}{2}(S^*(2\omega)-2 S(0))A^* (\tau)B(\tau)\, .
\end{eqnarray}
At this point it is worth to remark that the functions $A(\tau)$ and $B(\tau)$ that cancel the secular terms in $\mv{\vert Q(t)\vert^2}$ are different from those that cancel the secular terms in $\mv {Q(t)}$. This is of course due to the fact that
we are dealing with a stochastic differential equation, and therefore  $\vert\mv {Q(t)}\vert^2\neq\mv{\vert Q(t)\vert^2}$.

Note that the two differential Eqs.~\eqref{eq:diffAyB2} are enough to determine $\mv {|Q(t)|^2}$, that is given by
\begin{equation}\label{eq:modQ2gen}
    \mv {|Q(t)|^2}=|Y^{(0)}|^2+O(\epsilon^2)=|A(\tau)|^2+|B(\tau)|^2 + 2 \Re[A^*(\tau)B(\tau) e^{2 i \omega t}]+ O(\epsilon^2)\, .
\end{equation}
Using the adequate initial conditions $A(0)=1/\sqrt{2\omega}$, $B(0)=0$ we find for the Bogoliubov coefficient, 
\begin{equation}
    \mv{|\beta|^2}=\frac{1}{2}\left(e^{\omega^2\Re[S(2\omega)]\epsilon^2 t}-1\right)\, .
\end{equation}
This is the main result of this section: the stochastic mean value of the modulus squared of the Bogoliubov coefficient grows exponentially with a rate $\epsilon^2\omega^2\Re[S(2\omega)]$. 

As a test of our approach, we make contact with previous results in the literature. The procedure described above can be applied to compute $\mv{Q^2(t)}$ instead of $\mv{|Q(t)|^2}$. Doing this, with the initial conditions $Q(0)=1$ and $\dot Q(0)=0$ we obtain
\begin{eqnarray}\label{eq:modQ2Pap}
\mv{Q^2(t)}&=&
    \frac{1}{2} e^{\frac{\omega^2}{2} \left[ \Re[S(2\omega)] -2 S(0) \right] \epsilon^2 t} \cos\left[\left(2\omega- \frac{\omega^2 \epsilon^2}{2} \Im[S(2\omega)]\right)t \right] \nonumber\\
    &+& \frac{1}{2}
    e^{\omega^2  \Re[S(2\omega)]\epsilon^2 t}
    +O(\epsilon^2) \, ,
\end{eqnarray}
that coincides with the result obtained in Ref.~\cite{Papanicolau}.
Eq.~\eqref{eq:modQ2Pap} can be obtained from Eq.~\eqref{eq:modQ2gen} with the initial conditions $A(0)=B(0)=1/2$.

We can apply the analysis of this section to the  case of DCE in a cubic cavity. In the single-mode approximation, we set $g_{\mathbf{kj}}=0$ in Eq.~\eqref{eq:modes_perturbative}. The resulting equation is of the form of Eq. \eqref{eq:random_harmonic_oscillator}. Therefore,  the number of created particles  due to the stochastic motion of one of its walls reads
\begin{equation}\label{eq:Bog_single_stoc}
\mv{|\beta_\mk|^2}=\frac{1}{2}\left(e^{\frac{4\omega_{k_z}^4}{\omega_\mk^2}\Re[S(2\omega_\mk)]\epsilon^2 t}-1\right)\, .
\end{equation}
We see the different time scales in the rate of particle creation in the stochastic and deterministic situations: $\tau=\epsilon^2 t$ in the former and  $\bar\tau=\epsilon t$ in the latter. The MSA results are  compatible with the perturbative results discussed at the end  of Section 3: at  short times,  Eq.~\eqref{eq:Bog_single_stoc} is proportional to $\epsilon^2 t$, while from Eq.~\eqref{eq:Bog_single_det} we see that the square of modulus of the $\beta$ coefficient in the deterministic case is proportional to $\epsilon^2t^2$.

Although the single-mode approximation is a typical approximation in quantum optics, the perturbative calculations of Section 3
suggest that it is not justified for a noisy excitation of the system.  In the next section we will  relax the single mode assumption and consider the case in which there is intermode coupling.

\section{Nonperturbative calculation: coupled modes}

In order to assess the relevance of intermode couplings we will analyze the solutions of Eq. \eqref{eq:modes_perturbative} using MSA. As before, we will first compute the stochastic mean value $\mv{Q_{\mathbf{k}}^{(\mathbf{n})}}$ and then the Bogoliubov coefficients. In order to simplify the notation, we will omit the supraindex $(\mathbf{n})$
in the intermediate calculations.

Using the expansion
\begin{equation}
    Q_{\mathbf{k}}(t)= Y^{(0)}_{\mathbf{k}}(t,\tau) + \epsilon Y^{(1)}_{\mathbf{k}}(t,\tau) + \epsilon^2 Y^{(2)}_{\mathbf{k}}(t,\tau) + O(\epsilon^3)\, ,
\end{equation}
we obtain, from Eq.~\eqref{eq:modes_perturbative}, 
\begin{eqnarray}
\label{eq:rho_perturbation_equations_coupled}
   && \partial_t^2 Y^{(0)}_{\mathbf{k}} + \omega^2_{\mathbf{k}}Y^{(0)}_{\mathbf{k}}  = 0\, , \\
  && \label{eq:ordenl1} \partial_t^2 Y^{(1)}_{\mathbf{k}} + \omega^2_{\mathbf{k}} Y^{(1)}_{\mathbf{k}}  = 2\omega^2_{k_z} \xi(t) Y^{(0)}_{\mathbf{k}}
    + 2\dot\xi\sum_{\mathbf m} g_{\mathbf{k m}}\partial_t Y^{(0)}_{\mathbf{m}}
    +\ddot\xi\sum_{\mathbf m} g_{\mathbf{k m}}Y^{(0)}_{\mathbf{m}}\, ,\\
   && \label{eq:ordenl2}\partial_t^2 Y^{(2)}_{\mathbf{k}} + \omega^2_{\mathbf{k}} Y^{(2)} _{\mathbf{k}}  = - 2 \partial_{\tau t}^2 Y^{(0)}_{\mathbf{k}} 
    + 2\dot\xi\sum_{\mathbf m} g_{\mathbf{k m}}\partial_t Y^{(1)}_{\mathbf{m}}
    +\ddot\xi\sum_{\mathbf m} g_{\mathbf{k m}}Y^{(1)}_{\mathbf{m}} +\omega^2_{k_z}\xi Y^{(1)}_{\mathbf{k}}
    \, .
\end{eqnarray}
The zeroth order solution is
\begin{equation}
  Y^{(0)}_{\mathbf k}(t,\tau)=A_{\mathbf k}(\tau)e^{-i\omega_{\mathbf k}t} + B_{\mathbf k}(\tau)e^{i\omega_{\mathbf k}t},
\end{equation}
and the first and second order solutions $(l=1,2)$ are
\begin{equation}\label{eq:YJ}
  Y^{(l)}_{\mathbf k}(t,\tau) = \int_0^t dt'\, \frac{\sin[\omega_{\mathbf k}(t-t')]}{\omega_{\mathbf k}}J^{(l)}_{\mathbf k}(t',\tau)\, ,
  \end{equation}
  where we have introduced the notation $J^{(1)}_{\mathbf k}(t,\tau)$ and 
$J^{(2)}_{\mathbf k}(t,\tau)$ for the r.h.s. of Eqs.~\eqref{eq:ordenl1} and \eqref{eq:ordenl2}, respectively.

\subsection{MSA for coupled oscillators}

As before, the slowly varying functions  $A_{\mathbf k}(\tau)$ and $B_{\mathbf k}(\tau)$ are fixed by imposing that $\mv {Q_{\mathbf{k}}(t)}$ does not have secular terms, 
that is, $\mv{J^{(l)}_{\mathbf k}(t',\tau)}$ should not have terms proportional to $e^{\pm i\omega_{\mathbf k} t}$ (see Eq.~\eqref{eq:YJ}). 
Being linear in the noise, 
$\mv{J^{(1)}_{\mathbf k}(t',\tau)}$ vanishes; 
only $\mv{J^{(2)}_{\mathbf k}(t',\tau)}$ is relevant.

We illustrate the calculation with the first two terms in $\mv{J^{(2)}_{\mathbf k}(t',\tau)}$, that are $(a)\equiv-2\partial_{\tau t}^2 Y^{(0)}_{\mathbf{k}}(t',\tau) $ and $
    (b)\equiv 2\mv{\dot\xi\sum_{\mathbf m} g_{\mathbf{k m}}\partial_t Y^{(1)}_{\mathbf{m}}(t',\tau)}$. The first 
term does not have a stochastic contribution and reads
\begin{equation}
(a)=2 i \omega_{\mathbf k}\left[A'_{\mathbf k}e^{-i\omega_{\mathbf k}t} - B'_{\mathbf k}e^{i\omega_{\mathbf k}t}
\right]\, .
\end{equation}
The evaluation of the second term is more complex. We have
\begin{eqnarray}\label{eq:b}
(b)&=& 2 \sum_{\mathbf m} g_{\mathbf{k m}}\int_0^{t'} \cos[\omega_{\mathbf k}(t'-t'')]\Bigl\{ 2 \omega^2_{k_z}\mv{\dot\xi(t')\xi(t'')}Y_{\mathbf m}^{(0)}(t'',\tau)
 \nonumber\\
&& + 2\mv{\dot\xi(t')\dot\xi(t'')}\sum_{\mathbf n} g_{\mathbf{k n}}\partial_t Y^{(0)}_{\mathbf{n}}(t'',\tau)
+\mv{\dot\xi(t')\ddot\xi(t'')}\sum_{\mathbf n} g_{\mathbf{k n}}Y^{(0)}_{\mathbf{n}}(t'',\tau)\Bigr\}.
    \end{eqnarray}
As the noise correlation functions depend on $t'-t''$, the $t'$-integral in Eq.~\eqref{eq:b} is a sum of terms of the form
\begin{equation}
I_{\mathbf {n k}}=\int_0^{t'} du\, \cos(\omega_{\mathbf k}u)\left[ G(u) A_{\mathbf n}e^{-i\omega_{\mathbf n}(t'-u)} + G^*(u) B_{\mathbf n}e^{i\omega_{\mathbf n}(t'-u)}\right]\, ,
\end{equation}
where the function $G(u)$ has the information of the correlation functions. Therefore
\begin{equation}
2 I_{\mathbf{n k}}= A_{\mathbf {n }}[\tilde G(\omega_{\mathbf n}+\omega_{\mathbf k})
+ \tilde G(\omega_{\mathbf n}-\omega_{\mathbf k})]e^{-i\omega_{\mathbf n}t'} 
+ B_{\mathbf {n }}[\tilde G(\omega_{-\mathbf n}+\omega_{\mathbf k})
+ \tilde G(-\omega_{\mathbf n}-\omega_{\mathbf k})]e^{i\omega_{\mathbf n}t'}\, ,
\end{equation}
where we introduced the notation
\begin{equation}
\tilde G(\nu)=\int_0^\infty du G(u) e^{i u \nu}\, .
\end{equation}
From these expressions it is easy to recognize the secular terms, that will depend on the Fourier transform of the correlation functions evaluated at 
$\pm\omega_{\mathbf k} \pm \omega_{\mathbf n}$.

Following this procedure one can obtain the differential equations that determine the functions
$A_{\mathbf k}(\tau)$ and $B_{\mathbf k}(\tau)$. The result is
\begin{eqnarray}
&& A'_{\mathbf k}(\tau)+\lambda_{\mathbf k} A_{\mathbf k}(\tau)=0 \,,\nonumber\\
&& B'_{\mathbf k}(\tau)+\lambda_{\mathbf k} ^*B_{\mathbf k}(\tau)=0\, ,
\end{eqnarray}
with
\begin{equation}
\lambda_{\mathbf k}=\frac{\omega_{k_z}^4}{\omega_{\mathbf k}^2}(S(0)-S(2\omega_{\mathbf k}))-\sum_{\mathbf m}\frac{g^2_{\mathbf{k m}}}{4\omega_{\mathbf k}\omega_{\mathbf m}}(\omega_{\mathbf k}
^2-\omega_{\mathbf m}^2)^2
(S(\omega_{\mathbf k}+\omega_{\mathbf n})-S(\omega_{\mathbf k}-\omega_{\mathbf n}))\, .
\end{equation}
Note that, while the rate of change $\lambda_{\mathbf k}$ depends on the intermode coupling constants $g_{\mathbf{k m}}$, the differential  equations for the modes are uncoupled. This is somewhat unexpected. However, a closer look at the equations
reveals that this should be the case for a nondegenerate spectrum, because the MSA condition, i.e. the absence of secular terms in $\mv{J^{(2)}_{\mathbf k}(t',\tau)}$, is linear in the zeroth order solution
$Y^{(0)}_{\mathbf m}$. Therefore, the MSA condition selects the terms with $\mathbf {m}=\mathbf {k}$ if the spectrum is nondegenerate.

Restoring the supraindex $(\mathbf n)$, and taking into account the initial conditions,  the MSA solution for $\mv{Q_{\mathbf{k}}^{(\mathbf{n})}}$ reads
\begin{equation}
\mv{Q_{\mathbf{k}}^{(\mathbf{n})}}= \frac{\delta_{\mathbf{k,n}}}{\sqrt{2\omega_{\mathbf k}}}
e^{-i\omega_{\mathbf k} t-\epsilon^2 \lambda_{\mathbf k}t} + O(\epsilon^2)\, .
\end{equation}

\subsection{The Bogoliubov coefficients from MSA}

The calculation of 
the mean value of the modulus squared of the Bogoliubov coefficients is more complicated than in the case of uncoupled oscillators. The reason is the following. The generalization of Eq.~\eqref{eq:modQ2Bog} is, in this case
\begin{equation}
    |\nk {Q}|^2=\frac{1}{2\omega_\mk}\left(|\alpha_\mnk|^2+|\beta_\mnk|^2+2 \Re[\alpha_\mnk\beta_\mnk^*e^{-2i\omega_\mk t}]\right)\, .
\end{equation}
Taking into account that, in the MSA approximation
\begin{equation}
     \mv{|\nk {Q}|^2}=\left(|\nk{A}|^2+|\nk B|^2+2 \Re[\nk{A}
     B_\mk^{(\mn ) * }e^{-2i\omega_\mk t}]\right) + O(\epsilon^2)\, ,
\end{equation}
one can read the combination 
$\mv{|\alpha_\mnk|^2+|\beta_\mnk|^2}$ from the MSA solution.  The additional complication comes from the fact that, while for a single mode one has the condition $|\alpha|^2-|\beta|^2 =1$, for coupled oscillators 
the Bogoliubov coefficients satisfy
\begin{equation}
    \sum_{\mathbf k} \left(|\alpha_\mnk|^2-|\beta_\mnk|^2\right)=1\, .
\end{equation}
As a consequence, it is not possible to obtain both $|\alpha_\mnk|^2$and 
$|\beta_\mnk|^2$ from the MSA solution for $\mv{|\nk {Q}|^2}$ alone as in the single-mode case.  

To proceed, one can consider the additional
quantity 
\begin{equation}
    \nk{Q}\dot Q_\mk^{(\mn)*}=\frac{i}{2}\left(|\alpha_\mnk|^2-|\beta_\mnk|^2- 2 i \Im[\alpha_\mnk\beta_\mnk^*e^{-2i\omega_\mk t}]\right)\, ,
\end{equation}
and therefore compute $\mv{|\alpha_\mnk|^2-|\beta_\mnk|^2}$ from $\mv{\nk{Q}\dot Q_\mk^{(\mn)*}}$.
The procedure to determine $\mv{|\alpha_\mnk|^2}$ and $\mv{|\beta_\mnk|^2}$ is then to calculate both $\mv{\vert\nk{Q}\vert^2}$ and $\mv{\nk{Q}\dot Q_\mk^{(\mn)*}}$ using the MSA, imposing, in each case, the absence of secular terms. 

 Here we will sketch the computation of the {\it total} number of created particles, that can be obtained from $\mv{\vert\nk{Q}\vert^2}$.
Defining
\begin{equation}
    T_\mk^{(\mn)} (\tau)= |\nk{A}(\tau)|^2+|\nk B(\tau)|^2
 \, ,
\end{equation}
after a long calculation one can show that, to avoid the secular terms in 
$\mv{|\nk {Q}|^2}$,  $T_\mk^{(\mn)}$ must satisfy the differential equation
\begin{equation}\label{eq:diffT}
T_{\mk}^{(\mn )'}(\tau)+\gamma_\mk T_{\mk}^{(\mn )}(\tau) + \sum_{\mathbf m}\rho_{\mathbf{m k}}T_{\mm}^{(\mn )}(\tau)=0\, ,
\end{equation}
where
\begin{equation}
\gamma_\mk = -4\frac{\omega_{k_z}^4}{\omega_\mk^2}\Re[S(2\omega_\mk)]-\sum_\mm\left\{\frac{g_{\mathbf{k m}}^2}{2 \omega_\mk\omega_\mm}(\omega_\mk^2-\omega_\mm^2)^2\Re[S(\omega_\mk+\omega_\mm)-S(\omega_\mk-\omega_\mm)]\right\},
\end{equation}
and
\begin{eqnarray}
\rho_{\mathbf{m k}}&=&-\frac{g_{\mathbf{k m}}^2}{2\omega_\mk^2}\Bigl[(\omega_\mm^2-2\omega_\mm\omega_\mk-\omega_\mk^2)(\omega_\mm-\omega_\mk)^2\Re[S(\omega_\mm-\omega_\mk)]\nonumber \\
&&+  (\omega_\mm^2+2\omega_\mm\omega_\mk-\omega_\mk^2)(\omega_\mm + \omega_\mk)^2\Re[S(\omega_\mm +\omega_\mk)] \Bigr]\, .
\end{eqnarray}
This is a set of coupled differential equations for the functions $\nk{T}(\tau)$, that must be solved with the initial condition $\nk{T}(0)=\delta_{\mn\mk}/(2\omega_\mk)$
. From the solution one could compute the total number of created particles, taking into account that, for $t>T$, 
\begin{equation}
\sum_\mk 2\omega_\mk\nk{T}= 1+ 2\sum_\mk\mv{\vert\beta_{\mn\mk}\vert^2}\, ,
\end{equation}
and that the mean value of the number of created particles $\mv{\mathcal N}$ is
\begin{equation}
\mv{\mathcal N}=\sum_\mn\mv{\mathcal {N}_\mn}= \sum_{\mn\mk}\mv{\vert\beta_{\mn\mk}\vert^2}\, .
\end{equation}
It is worth to remark that, while in the deterministic case the mode coupling  depends on the spectrum of the cavity \cite{Crocce2001}, for a random motion it  occurs generically.

The solutions of
Eq.~\eqref{eq:diffT} will be described in a forthcoming work, as well
as the computation of the spectrum of created particles.

\section{Remarks on cosmological stochastic particle creation}
As already mentioned in the Introduction, SPC has also been considered in the context of quantum field theory in curved spacetimes, and in particular in a cosmological scenario. Assuming a flat Robertson-Walker metric
\begin{equation}\label{metric}
ds^2= a^2(\eta) (-d\eta^2+ d\mathbf x^2)\, ,
\end{equation}
where $\eta$ is the conformal time and $a(\eta)$
the scale factor of the Universe, the dynamical equation for the Fourier mode of momentum $k$ of a free quantum  scalar field $\phi$ is  \cite{booksQFTCS}
\begin{equation}
\phi_k'' + \left[k^2+m^2 a^2+(\chi-1/6) R a^2\right]\phi_k =0\, .
\end{equation}
Here a prime denotes derivative with respect to the conformal time,  $m$ is the mass of the field, $R$ the scalar curvature, and $\chi$ the coupling to the curvature. The metric  may have a stochastic component, as described in Ref.~\cite{Hu:1997iu}.

If the quantum field $\phi$ is coupled to another field $\varphi$ through $L_{int}=-(\lambda/2) \phi^2\varphi^2$, the equation for the $\phi_k$ mode reads
\begin{equation}
\phi_k'' + \left[k^2+m^2 a^2+(\chi-1/6) R a^2 + \lambda a^2\mv{\varphi^2}(t) + \lambda a^2 \xi_{\varphi}(t)\right]\phi_k =0\, .
\end{equation}
 This Langevin equation can be obtained formally by integrating out the field $\varphi$, in the context of effective field theories \cite{HuCal}. The equation contains a dissipative term proportional to $\mv {\varphi^2}$ and a multiplicative
 noise $\xi_\varphi$ that describes the fluctuations of $\varphi^2$ around its mean value. Note that 
the equations for the modes correspond to a set of uncoupled harmonic oscillators with  a time-dependent and stochastic frequency.  

From the discussion above we see that there are several sources of stochastic behaviour that may induce SPC for the quantum field. If the time scale of the stochastic fluctuations is much shorter than $H^{-1}$ ($H$ is the Hubble constant), the time dependence of the non-stochastic component of the metric can be neglected, and the equations for the Fourier modes are of the form
\begin{equation}\label{eq:modesapprox}
\phi_k'' + \left[k^2+M^2(1+\epsilon\xi(t)) \right]\phi_k =0\, ,
\end{equation}
where $M$ is a constant. We included a factor $\epsilon$ to make contact with our previous notation. This equation has been analyzed in Refs.~\cite{Amin:2015ftc, Choudhury:2018rjl}, using the analogy with the time-independent one dimensional Schroedinger equation
\begin{equation}\label{eq:Schroed}
    -\frac{1}{2m}\frac{d^2\psi}{dx^2}+V_r(x)\psi
= E \psi,
\end{equation}
for the wave function $\psi(x)$ of an electron inside a wire with random impurities, modelled by a random potential $V_r(x)$. 

It is usual to describe $V_r(x)$ as a set of potential barriers separated by a random distance, each barrier corresponding to a scatterer in the wire. Using the formalism of the transfer matrix, one can compute the reflection and transmission coefficients across a large number of scatterers, with the result \cite{Amin:2015ftc}
\begin{equation}
    T=e^{-\frac{L \gamma}{\Delta x} }\, ,
\end{equation}
where $L$ is the length of the wire, $\Delta x$ the mean distance between scatterers, and $\gamma$ a constant that depends on the shape of the barriers, the 
 "Lyapunov exponent". This the the phenomenon known as Anderson localization.
Using the analogy between Eqs.~\eqref{eq:modesapprox} and \eqref{eq:Schroed}, one can show that the random time-dependent mass in Eq.~\eqref{eq:modesapprox} produces an exponential SPC, with
\begin{equation}
   \mv{ \vert\beta_k\vert^2}\propto e^{\mu(k)\eta}\, ,
\end{equation}
where $\mu(k)$ is the $k$-dependent rate \cite{Amin:2015ftc, Choudhury:2018rjl}.

The results we found in Section IV can be applied {\it mutatis mutandis} to the cosmological problem. In particular, instead of computing the SPC rate through the reflection and transmission coefficients, one can compute the stochastic mean value $\mv{\vert\phi_k\vert^2}$ using the stochastic version of multiple scale analysis.  The result for the Bogoliubov coefficient $\beta_k$ associated to Eq.~\eqref{eq:modesapprox} is
\begin{equation}
    \mv{|\beta_k|^2}=\frac{1}{2}\left(e^{(k^2+M^2)\Re[S(2\sqrt{k^2+M^2})]\epsilon^2 \eta}-1\right)\, .
\end{equation}
This is an alternative approach to this problem, that highlights the relation between the SPC rate and the correlation function of the noise. Moreover, the stochastic MSA could be generalized to more complex situations  where, in addition to noise, there is a deterministic time dependence in the effective mass of the Fourier modes. This is the case for instance,  when the time dependence of the metric cannot be neglected, or when considering particle creation during reheating in the presence of quantum noise \cite{Zanchin}.


\section{Conclusions}

In this paper we analyzed the phenomenon of particle creation produced by stochastic external conditions. We focused mainly on the DCE, for the case of an electromagnetic cavity in which one of the mirrors has a stochastic motion. Physically, the motion of the mirror can be produced by thermal fluctuations, or due to the coupling to other system (environment). We have presented both perturbative and nonperturbative evaluations of the number of created particles. The perturbative approach is valid for short times due to the presence of secular terms. To analyze the evolution of the system for longer times, we adapted the MSA to the stochastic motion of the mirror. Although the fact that the stochastic terms can produce parametric resonance for a single harmonic oscillator with multiplicative noise is well known \cite{Papanicolau}, we have rederived those results using the conventional approach of MSA, and generalized the result to compute the Bogoliubov coefficients. We have also described the stochastic parametric resonance for the case of coupled harmonic oscillators. This is crucial for the DCE, since in general a noisy moving mirror will couple the different electromagnetic modes in the cavity. Unlike the deterministic case, where the mode coupling takes place only when  the spectrum of the cavity is such that the external frequency $\Omega$ satisfies $\Omega=|\omega_{\mathbf k}\pm\omega_{\mathbf j}\vert $ for some particular eigenfrequencies, the intermode coupling in the stochastic counterpart is  generic. This is indeed due to the fact that the random motion of the mirror can be thought as containing several frequencies. 

Although the number of created particles is in general much smaller in the stochastic than in the deterministic situation, one could in principle observe this kind of phenomena in superconducting cavities. The deterministic DCE
was first observed 
using a
transmission line terminated by a SQUID, by applying a time-dependent magnetic flux through it \cite{Wilson}. The SPC could be observed in this context by applying a noisy magnetic flux. The spectrum of created particles would be very different in both cases, in particular if the experiment is performed in a closed cavity.

We have also made contact with calculations of stochastic particle creation in a cosmological context. In this case, the different modes of the field are not coupled. Previous works in the subject emphasized  the connection with the resistance in one dimensional chains with impurities. Here we have shown that the same results can be interpreted as produced by stochastic parametric resonance.

\section*{Acknowledgments}
This research was supported by Agencia Nacional de Promoci\'on Cient\'ifica y Tecnol\'ogica (ANPCyT), Consejo Nacional de Investigaciones Cient\'ificas y T\'ecnicas (CONICET), and Universidad Nacional de Cuyo (UNCuyo). The work of L.G.T.~is supported by the Grant Agency of the Czech Republic, GACR grant 20-28525S.


\begin{thebibliography}{999}
\bibitem{reviews}
V.V.  Dodonov, Phys. Scr. {\bf 82}, 038105 (2010);  
D.A.R. Dalvit, P.A.  Maia Neto, and F.D.  Mazzitelli,  Lect. Notes Phys. {\bf 834}, 419 (2011);
P.D.  Nation, J.R.  Johansson, M.P.  Blencowe,  and F. Nori,  Rev. Mod. Phys. {\bf 84}, 1 (2012); 
V.V.  Dodonov, Physics {\bf 2}, 67 (2020).
\bibitem{booksQFTCS}
N.~D.~Birrell and P.~C.~W.~Davies,
``Quantum Fields in Curved Space,''
Cambridge Univ. Press, 1984;
E.~A.~Calzetta and B.~L.~B.~Hu,
``Nonequilibrium Quantum Field Theory,''
Cambridge University Press, 2008;
L.~E.~Parker and D.~Toms,
``Quantum Field Theory in Curved Spacetime: Quantized Field and Gravity,''
Cambridge University Press, 2009.
\bibitem{Hu:1997iu}
B.~L.~Hu and K.~Shiokawa,
Phys. Rev. D \textbf{57}, 3474-3483 (1998)
\bibitem{Amin:2015ftc}
M.~A.~Amin and D.~Baumann,
JCAP \textbf{02}, 045 (2016)
\bibitem{Choudhury:2018rjl}
S.~Choudhury, A.~Mukherjee, P.~Chauhan and S.~Bhattacherjee,
Eur. Phys. J. C \textbf{79}, no.4, 320 (2019)
\bibitem{Lozano:2020xga}
E.~Lozano and F.~D.~Mazzitelli,
Int. J. Mod. Phys. D \textbf{30}, no.15, 2150117 (2021)
\bibitem{Dodonov3}A.V.Dodonov, E.V.Dodonov, and V.V.Dodonov,
Phys. Lett.  A
\textbf{317}, 378(2003).
\bibitem{Nori} A. Settineri, V. Macri, L. Garziano, O. Di Stefano, F. Nori, and S. Savasta Phys. Rev. A \textbf{100}, 022501 (2019).
\bibitem{Roman}R. Roman-Ancheyta, I. Ramos-Prieto, A. Perez-Leija, K. Busch, and R. de J. Leon-Montiel
Phys. Rev. A \textbf{96}, 032501 (2017).
\bibitem{Moore} G.T. Moore, G. T.,  Journal of Mathematical Physics, {\bf 11}, 2679 (1970).
\bibitem{Crocce2001} 
M.~Crocce, D.~A.~R.~Dalvit and F.~D.~Mazzitelli,
Phys. Rev. A \textbf{64}, 013808 (2001).
\bibitem{Bender} C. M. Bender and S.A. Orszag, Advanced Mathematical Methods for Scientists and Engineers. McGraw-Hill, New York,(1978).
\bibitem{Hertz} 
M.~Crocce, D.~A.~R.~Dalvit, F.C. Lombardo and F.~D.~Mazzitelli, J. Opt. B: Quantum Semiclass. Opt. {\bf 7}, S32 (2005).
\bibitem{Ji} J.~Y.~Ji, H.~H.~Jung, J.~W.~Park and K.~S.~Soh,
Phys. Rev. A \textbf{56}, 4440 (1997).
\bibitem{Papanicolau} G. Papanicolau
and J. B. Keller, SIAM Journal on Applied Mathematics \textbf{21}, 287 (1971).
\bibitem{Wilson}
C.M. Wilson, G. Johansson, A. Pourkabirian, M. Simonen, J.R. Johansson et al.
Nature \textbf{479}, 376 (2011).
\bibitem{HuCal} See Calzetta and Hu, in Ref.~\cite{booksQFTCS}.
\bibitem{Zanchin} 
V.~Zanchin, A.~Maia, Jr., W.~Craig and R.~H.~Brandenberger,
Phys. Rev. D \textbf{57}, 4651, (1998).


\end{thebibliography}
\end{document}